\documentclass[prd,aps,secnumarabic,superscriptaddress,floatfix,preprintnumbers]{revtex4-1}
\usepackage[dvips]{graphicx,color}
\usepackage{dcolumn}
\usepackage{amsmath}
\def\mc{\multicolumn}

\def\st{\mbox{\rule[-3pt]{0pt}{14pt}}}

\def\Dot{\!\cdot\!}
\def\al{\alpha}
\def\be{\beta}
\def\ga{\gamma}
\def\de{\delta}

\def\ka{\kappa}
\def\la{\lambda}

\def\sig{\sigma}
\def\ep{\varepsilon}
\def\part{\partial}
\def\arccosh{\mbox{\rm arccosh}}
\def\arcsinh{\mbox{\rm arcsinh}}

\def\f#1#2{{\textstyle\frac{#1}{#2}}}
\def\ff#1#2#3{{\textstyle\frac{#1}{#2#3}}}
\def\arcsinh{{\rm arcsinh}}
\def\Li{{\rm Li}}
\def\Re{\mbox{\rm Re}}

\newcommand{\mathsym}[1]{{}}
\newcommand{\unicode}[1]{{}}

\begin{document}

\title{Muonic hydrogen and the proton size}

\author{Wayne W. Repko}\email{repko@pa.msu.edu}
\affiliation{Department of Physics and Astronomy, Michigan State University, East Lansing, MI 48824}
\author{Duane A. Dicus}\email{dicus@physics.utexas.edu}
\affiliation{Department of Physics, University of Texas, Austin, TX 78712}

\date{\today}
\begin{abstract}
We reexamine the structure of the $n=2$ levels of muonic hydrogen using a two-body potential that includes all relativistic, recoil and one loop corrections. The potential was originally derived from QED to describe the muonium atom and accounts for all contributions to order $\al^5$. Since one loop corrections are included, the anomalous magnetic moment contributions of the muon can be identified and replaced by the proton anomalous magnetic moment to describe muonic hydrogen with a point-like proton. This serves as a convenient starting point to include the dominant electron vacuum polarization corrections to the spectrum and extract the proton's mean squared radius $r_p=\sqrt{\langle r^2\rangle}$. Our results are consistent with other theoretical calculations that find that the muonic hydrogen value for $r_p$ is smaller than the result obtained from electron scattering.
\end{abstract}

\maketitle

\section{Introduction}

The muonic hydrogen experiments \cite{nature,science} measure both the $2\,^3S_{1/2}\leftrightarrow 2\,^5P_{3/2}$ and the $2\,^1S_{1/2}\leftrightarrow 2\,^3P_{3/2}$ transitions. The experimental results are, respectively, 49881.35(65) GHz = 206.2925(3) meV and 54611.16(1.05) GHz = 225.8535(4) meV. These measurements have been compared with a variety of theoretical calculations \cite{Borie,Borie_1,Pachucki,Pachucki_1,Pohl,Antognini,Peset} that include a dependence on the mean squared proton radius $\langle r^2\rangle$. Our purpose here is to compare the contributions to the Lamb shift that are independent of the proton structure with those of previous calculations. We find a value of $205.980\;{\rm meV}$, about $0.07\;{\rm meV}$ smaller than other theoretical calculations. 

If proton structure corrections are included, the resulting values of $r_p$ from muonic hydrogen are systematically smaller than those generally obtained from electron scattering data \cite{MTN}, leading to a disparity between the two approaches. Some of disparity could be associated with uncertainties in the scattering data. The proton radius experiment (PRad) at the Jefferson Laboratory is designed to address this issue \cite{jlab}.  A recent spectroscopic measurement of the Rydberg constant \cite{Beyer} reports a smaller value of $r_p$ consistent with muonic hydrogen, but a new spectroscopic measurement of the $1S\to 3S$ transition in hydrogen \cite{fleur} supports a larger value as do most other spectroscopic measurements.

Here, we reexamine the theoretical calculation from a slightly different starting point. Our approach is to modify the two-body potential originally derived from QED to describe the muonium atom \cite{GRS}. This potential contains all relativistic, recoil and one-loop terms that contribute to order $\al^5$. The inclusion of the one-loop corrections enables us to identify the muon anomalous magnetic moment and replace it by the proton's anomalous magnetic moment $\ka=1.79285$. The resulting potential can be used to calculate the fine structure, hyperfine structure, Lamb shift and recoil corrections for muonic hydrogen with a a point-like proton. It also serves as the starting point to include the dominant electron vacuum polarization contributions. The resulting hyperfine, spin-orbit, tensor and spin-independent potentials are \cite{SNGSFR}
\begin{eqnarray}
V_{HF} &=& \frac{4\pi\al}{m_1m_2}\left[\frac{2}{3}(1+a_\mu)(1+\ka)+ \frac{\al}{\pi}\frac{m_1m_2}{m_1^2-m_2^2}\ln\left(\frac{m_2^2}{m_1^2}\right)\right] \vec{S}_1\Dot\vec{S}_2\,\de(\vec{r})\,,\label{VHF}\\[6pt]
V_{LS} &=& \frac{\al}{r^3}\left[\frac{(1+2a_\mu)}{2m_1^2}+\frac{(1+a_\mu)}{m_1m_2}\right]\vec{L}\Dot\vec{S}_1+ \frac{\al}{r^3}\left[\frac{(1+2\ka)}{2m_2^2}+\frac{(1+\ka)}{m_1m_2}\right]\vec{L}\Dot\vec{S}_2 \,,\label{VLS}\\[6pt]
V_T &=& \frac{\al(1+a_\mu)(1+\ka)}{m_1m_2r^3} \left(3\vec{S}_1\Dot\hat{r}\vec{S}_2\Dot\hat{r}-\vec{S}_1\Dot\vec{S}_2\right)\,,
\label{VT} \\[6pt]
V_{SI} &=& -\frac{\al}{m_1m_2r}\vec{p}^{\;2}+\frac{\al\pi}{2\mu^2}\de(\vec{r})-\frac{1}{8} \left(\frac{1}{m_1^3}+\frac{1}{m_2^3}\right) (\vec{p}^{\;2})^2+\frac{\al^2\mu}{2m_1m_2r^2} \nonumber \\ [6pt]
&& +\al^2\left[\frac{4}{3}\left(\frac{1}{\mu^2}\ln(\frac{\mu}{\la_{IF}})-\frac{1}{m_1^2}\ln(\eta_2)- \frac{1}{m_2^2}\ln(\eta_1)\right)\de(\vec{r})+\left(\frac{2(m_1^2\ln(\eta_1)-m_2^2\ln(\eta_2))} {m_1m_2(m_1^2-m_2^2)}\right)\de(\vec{r})\right.\nonumber \\ [6pt]
&&\left. +\frac{7}{6\pi m_1m_2}\nabla^2\left(\frac{\ln(\mu r)+\ga}{r}\right)-\frac{4}{15}\left(\frac{1}{m_1^2}+\frac{1}{m_2^2}\right)\de(\vec{r})\right]\,.
\label{VSI}
\end{eqnarray}
Here, $\mu$ is the reduced mass, $\eta_i=m_i/(m_1+m_2)$, $\ga$ is Euler's constant and $\la_{IF}$ is the infrared cutoff. The last term in Eq.\,(\ref{VSI}) is the contribution from the muon and proton vacuum polarizations. In what follows, we use \cite{pdg} $m_1=105.6583715(35)\;{\rm MeV}$, $m_2=938.272046(21)\;{\rm MeV}$, $\mu=94.9645\;{\rm MeV}$, $m_e=0.510998928(11)\;{\rm MeV}$, $\al=1/137.035999074(44)$ and $a_\mu=\al/2\pi$.

Much of the reason for undertaking the following calculation is that Eq.\,(\ref{VSI}) contains several terms that differ from those commonly used in determining the muonic hydrogen spectrum. In particular, the order $\al$ delta function term behaves as $\mu^{-2}$ whereas the corresponding term in \cite{Pachucki} has an overall $(1/m_1^2+1/m_2^2)$ factor. The difference in the s-state contribution is several meV in the order $\al^4$ correction. Also, the one-loop term $\mu\,\al^2/(2m_1m_2r^2)$ contributes at order $\al^4$ at the few meV level. The $\mu^{-2}$ dependence of the order $\al^2$ one-loop $\ln(\mu/\la_{IF})$ term results in a recoil correction to the Lamb shift of order $\mu^3\,\al^5/m_2^2$ that is larger than the order $\al^6$ correction to this contribution given in \cite{Pachucki}. We undertook the calculation to determine the implication of these differences.

\section{Electron vacuum polarization effects}
\subsection{One loop correction to the Coulomb potential}
The dominant contributions to the $2P-2S$ splitting in muonic hydrogen are due to the electron vacuum polarization corrections to the photon propagator. These contributions can be included by using the dispersion representation for the photon propagator \cite{GK,HL,KS}
\begin{equation}\label{prop}
D(k^2)=\frac{1}{k^2}-\int_0^\infty\!\!\frac{d\la}{\la}\frac{\Delta(\la)}{\la-k^2}\,,
\end{equation}
where $\Delta(q^2)$ is
\begin{equation}
\Delta(q^2)=\frac{(2\pi)^3}{3q^2}\sum_n\de^{(4)}(q-q_n)\langle0|j_\mu(0)|n\rangle \langle n|j^\mu(0)|0\rangle\,.
\end{equation}
For the $e\,\bar{e}$ intermediate state, $\Delta^{(2)}(\la)$ is
\begin{equation}
\Delta^{(2)}(\la)=\frac{\al}{3\pi}(1+2m_e^2/\la)\sqrt{1-4m_e^2/\la}\;\theta(\la-4m_e^2)\,.
\end{equation}
If we take $k^2$ to be space-like, then the modified Coulomb interaction at the one-loop level is
\begin{eqnarray}
V(\vec{k}^{\,2}) &=&-\frac{e^2}{\vec{k}^{\,2}}-e^2\!\!\int_{4m_e^2}^\infty\!\frac{d\la}{\la} \frac{\Delta^{(2)}(\la)}{\vec{k}^{\,2}+\la} \\
&=& V_C(\vec{k}^{\,2})+V_{VP}(\vec{k}^{\,2})\,.
\end{eqnarray}
The explicit form of $V_{VP}(\vec{k}^{\,2})$ in momentum space can be obtained by integrating over $\la$ which results in
\begin{equation}
V_{VP}(\vec{k}^{\,2})=\frac{-e^2}{\vec{k}^{\,2}}\Pi^{(2)}_f(\vec{k}^{\,2})\,,
\end{equation}
with
\begin{equation}
.\Pi^{(2)}_f(\vec{k}^{\,2})= \frac{2\al}{\pi}\left[\frac{1}{3}(1-2m_e^2/\vec{k}^2) \left(\sqrt{1+4m_e^2/\vec{k}^2}\;\arcsinh\left(\frac{|\vec{k}|}{2m_e}\right) -1\right)+\frac{1}{18}\right]\,
\end{equation}
$\Pi^{(2)}_f(\vec{k}^{\,2})$ is the electron one-loop vacuum polarization correction in the spacelike region.

Transforming to coordinate space
\begin{eqnarray}
V_{VP}(\vec{r})&=&\frac{1}{(2\pi)^3}\int\!\!d^{\,3}k\,V_{VP}(\vec{k})e^{i\vec{k}\cdot\vec{r}}\\ [6pt]
V_{VP}(\vec{r})&=&-\frac{\al}{r} \int_{4m_e^2}^\infty\!\frac{d\la} {\la}\Delta^{(2)}(\la)e^{-\sqrt{\la}\,r}\,.
\end{eqnarray}
To compute the effect of $V_{VP}$, we need to calculate the difference between $\langle 21m|V_{VP}|21m\rangle $ and $\langle 200|V_{VP}|200\rangle$. For example, $\langle 21m|V_{VP}|21m\rangle$ is
\begin{eqnarray}
\langle 21m|V_{VP}|21m\rangle &=& -\frac{\al}{24 a}\int_{4m_e^2}^\infty\! \frac{d\la}{\la}\Delta^{(2)}(\la)\int_0^\infty\!\!dt\,t^3\,e^{-(1+a\sqrt{\la})t} \nonumber \\ [6pt]
\langle 21m|V_{VP}|21m\rangle &=& -\frac{\al}{4 a}\int_{4}^\infty\! \frac{dx}{x}\frac{\Delta^{(2)}(x)}{(1+m_ea\sqrt{x})^4}\label{2P}\,,
\end{eqnarray}
with $a=1/\mu\al$ and $\la=m_e^2x$. A similar calculation gives
\begin{equation}
\langle 200|V_{VP}|200\rangle = -\frac{\al}{4 a}\int_{4}^\infty\! \frac{dx}{x}\frac{\Delta^{(2)}(x)(1+2m_e^2a^2x)}{(1+m_ea\sqrt{x})^4}\label{2S}\,.
\end{equation}
The difference between Eqs.\,(\ref{2P}) and (\ref{2S}) is
\begin{equation}\label{2P-2S}
EE_{VP}=\Delta(2P-2S) = \frac{\mu\al^3}{6\pi}\int_{4}^\infty\! \frac{dx\,(1+2/x)\sqrt{1-4/x}\;m_e^2a^2}{(1+m_ea\sqrt{x})^4}=205.007\, {\rm meV}\,.
\end{equation}
This agrees with \cite{Pachucki}\,. It is worth noting that the muon contribution from Eq.\,(\ref{2P-2S}) is $0.0167$ meV which is virtually the same as the $0.0168$ meV result coming from $m_1^{-2}$ muon contribution in last term of Eq.\,(\ref{VSI}). The proton vacuum contribution from the $m_2^{-2}$ term in Eq.\,(\ref{VSI}) is indistinguishable from the result obtained using Eq.\,(\ref{2P-2S}).

\subsection{Two-loop correction to the Coulomb potential}

The two-loop contributions to $\Delta(\la)$, $\Delta^{(4)}(\la)$, have been calculated by K\"allen and Sabry \cite{KS}. They contain both the reducible double electron bubble diagram and the irreducible fourth order term $\Pi^{(4)}_f(\vec{k}^{\,2})$. These corrections can be expressed as a correction to the Coulomb potential in the form \cite{blom}
\begin{equation}
V_{VP2}(r) = -\frac{\al}{r}F(r)\,,
\end{equation}
where
\begin{eqnarray}\label{F(r)}
F(r) &=& -\frac{\al^2}{\pi^2}\int_1^\infty\!\!dt\,e^{-2tr}\Big[\left(\frac{13}{54 t^2}+\frac{7}{108 t^4}+\frac{2}{9 t^6}\right) \sqrt{t^2-1}+\left(\frac{4}{3 t^2}+\frac{2}{3 t^4}\right) \sqrt{t^2-1} \log\left(8t \left(t^2-1\right)\right) \nonumber \\ [6pt]
&&+\left(-\frac{44}{9 t}+\frac{2}{3 t^3}+\frac{5}{4 t^5}+\frac{2}{9 t^7}\right) \arccosh(t)+\left(-\frac{8}{3t}+\frac{2}{3
t^5}\right)\int_t^\infty\!\!dx\,f(x)\Big]\,,
\end{eqnarray}
with
\begin{equation}
f(x)=\frac{3x^2-1}{x(x^2-1)}\arccosh(x)- \frac{1}{\sqrt{x^2-1}}\log\left(8x(x^2-1)\right)\,.
\end{equation}
By transforming this potential to momentum space and comparing it with Eq.\,(\ref{prop}), $\la$ can be identified with $4t^2$ and the second order energy shift expressed as
\begin{equation}
EE_{VP2}=-\frac{\mu\al^4}{4\pi^2}\int_4^\infty\!\!dx\, \frac{\Delta^{(4)}(\sqrt{x}/2)\,m_e^2a^2}{(1+m_ea\sqrt{x})^4}=1.508\;{\rm meV}\,,
\end{equation}
where $\Delta^{(4)}(x)$ is
\begin{eqnarray}
\Delta^{(4)}(x) &=& \left(\frac{13}{54 x}+\frac{7}{108 x^3}+\frac{2}{9 x^5}\right) \sqrt{x^2-1}+\left(\frac{4}{3 x}+\frac{2}{3 x^3}\right) \sqrt{x^2-1} \log\left(8x \left(x^2-1\right)\right) \nonumber\\ [6pt]
&&+\left(-\frac{44}{9}+\frac{2}{3 x^2}+\frac{5}{4 x^4}+\frac{2}{9 x^6}\right) \arccosh(x)+\left(-\frac{8}{3}+\frac{2}{3
x^4}\right)\Bigg[\frac{2 \pi ^2}{3} \nonumber\\ [6pt]
&&-\arccosh^2(x)-\log\left(8x(x^2-1)\right)\arccosh(x)-2\Re[\Li_2\left((x+\sqrt{x^2-1})^2\right)] \nonumber \\ [2pt]
&&+\Li_2\left(-(x-\sqrt{x^2-1})^2\right)\Bigg]\,.
\end{eqnarray}
The terms in the large square brackets result from evaluating the last integral in Eq.\,(\ref{F(r)}) and $\Li_2(z)$ is the Spence function
$$
\Li_2(z)=-\int_0^1\!\!\frac{dt}{t}\log(1-zt)\,.
$$
\subsection{Three-loop correction to the Coulomb potential}
The three-loop contribution to $\Delta(\la)$, $\Delta^{(6)}(\la)$, is not available in the literature. These contributions have been calculated by Kinoshita and Nio \cite{KN}. They note that the three-loop correction is comprised of two reducible diagrams and one irreducible diagram that can be represented as
\begin{equation}
\left(\Pi^{(2)}_f(\vec{k}^{\,2})\right)^3+ 2\Pi^{(2)}_f(\vec{k}^{\,2})\Pi^{(4)}_f(\vec{k}^{\,2})+\Pi^{(6)}_f(\vec{k}^{\,2})\,.
\end{equation}
Remembering the reducible fourth order dispersion result
\begin{equation}
-e^2\!\!\int_{4m_e^2}^\infty\!\frac{d\la}{\la} \frac{\Delta^{(4)}(\la)}{\vec{k}^{\,2}+\la}=-\frac{e^2}{\vec{k}^{\,2}}\Pi^{(4)}(\vec{k}^{\,2})\,,
\end{equation}
and
\begin{equation}
\Pi^{(4)}(\vec{k}^{\,2})=\left(\Pi^{(2)}_f(\vec{k}^{\,2})\right)^2+\Pi^{(4)}_f(\vec{k}^{\,2})\,,
\end{equation}
it is possible to verify the contributions of the reducible terms. They are $0.000396$ meV and $0.0029312$ meV, respectively. The numerical evaluation of the $\Pi^{(6)}_f(\vec{k}^{\,2})$ contribution is \cite{KN} $.001103$ meV, giving a total three-loop correction to the Coulomb potential of $0.004431$ meV.

\subsection{$\al^4$ Second order non-relativistic perturbation correction}

The large size of the one-loop correction suggests that the contribution of $V_{VP}(\vec{r})$ in second order non-relativistic perturbation theory is not negligible. Evaluating this correction necessitates using the radial portions of the Coulomb Green's function for $n=2$ and $\ell=0,1$, expressed as $\mu^2\al\,g_{2\ell}(x,x')$. General expressions for these Green's functions were derived by Hostler \cite{HP,Hostler} and explicit expressions for small values of $n$ are contained in \cite{LF,JH}. Due to some typographical errors in the latter papers ($g_{20}$ in Ref.\,\cite{LF} and Eq.\,(2.18) in Ref.\,\cite{JH}), the expressions for $g_{20}(x,x')$ and $g_{21}(x,x')$ ($x=r/a,\;x'=r'/a$) are given here. 
\begin{subequations}
\begin{eqnarray}
g_{20}(x,x') &=& e^{-(x+x')/2}\left[(2-x)(2-x')\left(\ln(x)+\ln(x')+\frac{(x+x')}{4}+2\ga-\frac{15}{4}+{\rm Ei}(x_<)\right)\right. \nonumber \\
&& \left.+12-2x-2x'-\frac{2}{x}-\frac{2}{x'}+\frac{x}{x'}+\frac{x'}{x}-xx' +(2-x_>)e^{x_<}\left(\frac{1}{x_<}-1\right)\right]\frac{1}{4\pi}\,, \\ [6pt]
g_{21}(x,x') &=& \frac{xx'}{3}e^{-(x+x')/2}\left[\ln(x)+\ln(x')+ \frac{(x+x')}{4} +2\ga-\frac{49}{12}-\frac{3}{x}-\frac{3}{x^2}-\frac{2}{x^3} \right.\nonumber \\ 
&&\left.-\frac{3}{x'}-\frac{3}{x'^{\,2}}-\frac{2}{x'^{\,3}}-{\rm Ei}(x_<) +\left(\frac{1}{x_<}+\frac{1}{x_<^2} +\frac{2}{x_<^3}\right)e^{x_<} \right]\frac{3\hat{x}\Dot\hat{x}'}{4\pi}\,. 
\end{eqnarray}
\end{subequations}
Here, Ei(x) is
\begin{equation}
{\rm Ei}(x)=\int_{-\infty}^x\!\!dt\frac{e^t}{t}\,,
\end{equation}
and
\begin{equation}
x_<=x\theta(x'-x)+x'\theta(x-x')\qquad x_>=x'\theta(x'-x)+x\theta(x-x')\,.
\end{equation}
The contributions take the form
\begin{equation}
\mu^2\al\int_0^\infty\!\!drr^2R_{n\ell}(r)V_{VP}(r)\int_0^\infty\!\!dr'r'^{\,2}g_{n\ell}(x,x') V_{VP}(r')R_{n\ell}(r')\,,
\end{equation}
for $n=2$ and $\ell=0,1$. The integrations over $r$, $r'$ can be evaluated exactly using Mathematica, and integrations over propagator parameters $\la$ and $\la'$ can then be calculated numerically. The results are $EE^{(2)}2P_{VP}=-0.0022671$ meV and $EE^{(2)}2S_{VP}=-0.153164$ meV for a net contribution of \cite{Pachucki}
\begin{equation}
EE^{(2)}_{VP}=0.1509\,{\rm meV}\,.
\end{equation}
\subsection{$\al^5$ Second order non-relativistic perturbation correction}

There is also a second order non-relativistic perturbative contribution from the combination of a one loop vacuum polarization correction and a two loop vacuum polarization correction . The calculation is similar to the one loop second order calculation and results in a 0.00215 meV contribution. In addition, there is a third order non-relativistic perturbative contribution from three one-loop vacuum polarization corrections \cite{KN,Ivanov} which gives $0.00007$ meV.

\section{Proton size correction}

The effect of the proton size can be obtained in terms of its mean square radius $\langle r^2\rangle$ by modifying the Coulomb potential with a charge form factor $F(\vec{k}^{\;2})$ defined as
\begin{equation}\label{ff}
F(\vec{k}^{\;2})=\int\!\! d^3r\;\rho(r)\,e^{-i\vec{k}\cdot\vec{r}}\,,
\end{equation}
where $\rho(r)$ is the proton charge density. In momentum space, the Coulomb potential becomes
\begin{equation}
V(\vec{k}^{\;2})=-\frac{e^2 F(\vec{k}^{\;2})}{\vec{k}^{\;2}}\,.
\end{equation}
Expanding the exponential in Eq.\,(\ref{ff}) and integrating gives
\begin{equation}
F(\vec{k}^{\;2})\sim 1-\frac{1}{6}\langle r^2\rangle\vec{k}^{\;2}+\cdots\,,
\end{equation}
so $V(\vec{q}^{\;2})$ is approximately
\begin{equation}
V(\vec{k}^{\;2})\sim -\frac{e^2}{\vec{k}^{\;2}}+\frac{e^2}{6}\langle r^2\rangle\,.
\end{equation}
This gives
\begin{equation}
V(r)\sim -\frac{\al}{r}+\frac{2}{3}\pi\al \langle r^2\rangle\,\de(\vec{r})\,.
\end{equation}
The perturbative contribution due to a finite proton radius is then
\begin{equation}
E(\langle r^2\rangle)=\frac{2}{3}\frac{\mu^3\al^4\langle r^2\rangle}{n^3}\de_{\ell\,0}\,.
\end{equation}
Numerically for $n=2$, this is 
\begin{equation}\label{<r^2>}
E(\langle r^2\rangle)=5.1975\,\langle r^2\rangle\;{\rm meV/fm^2}.
\end{equation}
Since the reported energy difference is $P_{3/2}-S_{1/2}$, the sign of the contribution is negative. The electron vacuum polarization corrections to Eq.\,(\ref{<r^2>}) are calculated in the Appendix. We do not directly address the order $\langle r^3\rangle/a^3$ proton size correction, but any single parameter functional form for $F(\vec{k}^{\;2})$ that satisfies $\vec{k}^{\,2}dF(\vec{k}^{\;2})/d\vec{k}^{\,2}=-\langle r^2\rangle/6$ can give an estimate of the size of this correction. For example, if $F(\vec{k}^{\;2}) = (r_p^2k^2/12+1)^{-2}$ the correction to the Coulomb interaction is
\begin{equation}
\Delta V(r)=\frac{\al}{r}(1+ \frac{\sqrt{3}\,r}{r_p}) e^{-2\sqrt{3}\,r/r_p}\,,\,
\end{equation}
and
\begin{equation}
\langle 2S|\Delta V(r)|2S\rangle\simeq \mu\al^2 \left(\frac{r_p^2}{12\,a^2}-\frac{5r_p^3}{24\sqrt{3}\,a^3}+\cdots\right)= 5.1975\,r_p^2\;{\rm meV\,fm^{-2}}-0.0263\,r_p^3\;{\rm meV\,fm^{-3}}\,.
\end{equation}
The ${\cal O}(r_p^3/a^3)$ coefficient is roughly the size obtained by more detailed calculations \cite{MF1,MF2,PH,GP}. The $2P$ contribution is ${\cal O}(r_p^4/a^4)$.

The remaining corrections to the energy levels come from the potentials in Eqs.\,(\ref{VHF}-\ref{VSI}). In what follows, we use simultaneous eigenstates of $\vec{F}^{\,2},F_z,\vec{J}^{\,2},\vec{S_2}^{\,2}$, where $\vec{F}=\vec{L}+\vec{S}_1+\vec{S}_2$, $\vec{J}=\vec{L}+\vec{S}_1$.

\section{Spin Independent Terms}
\subsection{Order $\al^4$ terms}
The terms in the first line of Eq.\,(\ref{VSI}) contribute to the $s$ and $p$ levels in order $\al^4$. Their expectation values are (in meV)
\begin{subequations}
\begin{eqnarray}
\langle -\frac{\al}{m_1m_2r}\vec{p}^{\;2}\rangle &=& -\frac{\mu^3\al^4}{n^3 m_1 m_2}\left(\frac{4}{2\ell + 1} - \frac{1}{n}\right) 10^9 \\ [6pt]
\langle \frac{\al\pi}{2\mu^2}\de(\vec{r})\rangle &=& \frac{\mu\al^4}{2n^3}\de_{\ell 0}\,10^9 \\ [6pt]
\langle -\frac{1}{8} \left(\frac{1}{m_1^3}+\frac{1}{m_2^3}\right) (\vec{p}^{\;2})^2\rangle &=& -\frac{\mu\al^4}{8n^3}\left(\frac{\mu^3}{m_1^3} +\frac{\mu^3}{m_2^3}\right)\left(\frac{8}{2\ell +1}-\frac{3}{n}\right)10^9 \\ [6pt]
\langle \frac{\al^2\mu}{2m_1m_2r^2}\rangle &=& \frac{\mu^3\al^4}{n^3m_1m_2} \frac{1}{2\ell +1}10^9
\end{eqnarray}
\end{subequations}
and it should be noted that the $1/r^2$ term is part of the one-loop correction. For the $n=2$ state, the contributions are
\begin{subequations}
\begin{eqnarray}
E_2(2S) &=& -10.711\,{\rm meV} \\
E_2(2P) &=& -5.100\, {\rm meV} \\
E_2 &=& E_2(2P)-E_2(2S)=5.610\, {\rm meV}
\end{eqnarray}
\end{subequations}
\subsection{Order $\al^5$ terms}
The remainder of the terms in Eq.\,(\ref{VSI}) are of order $\al^5$ or $\al^5\ln(\al)$. There are two issues to address when evaluating these terms. The first is the elimination of the photon mass dependence. This is accomplished by using the `Bethe logarithm' technique, which amounts to the replacement of $\ln(\mu/\la_{IF})$ by \cite{EY}
\begin{equation}
\left(\ln\frac{R_\infty}{\al^2k_0(n,0)}+\frac{5}{6}\right)\de_{\ell 0}+ \ln\frac{R_\infty}{k_0(n,\ell)}(1-\de_{\ell 0})\,.
\end{equation}
The other is the matrix element of $\nabla^2[(\ln(\mu r)+\ga)/r]$. For states with $\ell>0$, this reduces to $-1/r^3$. When $\ell=0$, the result is
\begin{equation}
\langle n0|\,\frac{1}{2\pi}\nabla^2\left[\frac{\ln(\mu r)+\ga}{r}\right]|n0\rangle=\frac{2\mu^3\al^3}{\pi n^3}\left[\ln\frac{2\al}{n} +\frac{n-1}{2n}+\sum_{k=1}^n\,\frac{1}{k}\right]\,.
\end{equation}
Using
\begin{eqnarray}
\ln k_0(2S) &=& 2.8117699 \\
\ln k_0(2P) &=& -0.0300167
\end{eqnarray}
and denoting the expectation values of the order $\al^5$ as $E_4(n\ell)$, the results are
\begin{subequations}
\begin{eqnarray}
E_4(2S) &=& 0.7077\,{\rm meV} \\
E_4(2P) &=& 0.0004\, {\rm meV} \\
E_4 &=& E_4(2P)-E_4(2S)= -0.7073\, {\rm meV}
\end{eqnarray}
\end{subequations}

\section{Spin dependent terms}
\subsection{$\ell=0$ Hyperfine}
The expectation value of $V_{HF}$ affects only $s$-states and is
\begin{subequations}
\begin{eqnarray}
E_{HF}(nS_{1/2})&=&\frac{\mu^3\al^4}{n^3m_1m_2}\left[\frac{2}{3}(1+a_\mu)(1+\ka)+ \frac{\al}{\pi}\frac{m_1m_2}{m_1^2-m_2^2}\ln\left(\frac{m_2^2}{m_1^2}\right)\right]
(2s(s+1)-3)\\ [6pt]
E_{HF}(^3S_{1/2}) &=& 5.704\,{\rm meV} \\ [4pt]
E_{HF}(^1S_{1/2}) &=& -17.113\,{\rm meV} \\ [4pt]
\Delta E_{HF}(S_{1/2}) &=& 22.817\,{\rm meV}
\end{eqnarray}
\end{subequations}
\subsection{Spin-orbit and Tensor terms}
The largest contribution of $V_{LS}$ is that associated with the $\vec{L}\Dot\vec{S}_1$ term. It accounts for the fine structure splitting between the $P_{3/2}$ and the $P_{1/2}$ states. The contribution is
\begin{equation}
E_{FS}(P_{j_1})=\al\left[\frac{(1+2a_\mu)}{2m_1^2}+\frac{(1+a_\mu)}{m_1m_2}\right] \langle\frac{1}{r^3}\rangle\,\langle \vec{L}\Dot\vec{S}_1\rangle\,.
\end{equation}
The expectation value of $r^{-3}$ is
\begin{equation}
\langle\frac{1}{r^3}\rangle=\frac{2\mu^3\al^3}{n^3\ell (\ell +1)(2\ell +1)}\to\frac{\mu^3\al^3}{24}\,.
\end{equation}
Since the eigenstates we are using are eigenstates of $\vec{J}^{\,2}$,
\begin{equation}
2\vec{L}\Dot\vec{S}_1=\vec{J}^{\,2}-\vec{L}^{\,2}-\vec{S}_1^{\,2}=j(j+1)-2-3/4\,.
\end{equation}
Then,
\begin{equation}
E_{FS}(P_{j})=\frac{\mu^3\al^4}{48m_1^2}\left[(1+2a_\mu)+ \frac{2m_1}{m_2}(1+a_\mu)\right]\langle \vec{L}\Dot\vec{S}_1\rangle\,,
\end{equation}
and
\begin{subequations}
\begin{eqnarray}
E_{FS}(P_{3/2}) &=& 2.782\,{\rm meV}\,, \\
E_{FS}(P_{1/2}) &=& -5.564\, {\rm meV}\,, \\
\Delta E_{FS} &=& E_{FS}(P_{3/2})-E_{FS}(P_{1/2})=8.347\, {\rm meV}\,.
\end{eqnarray}
\end{subequations}

The remaining spin dependent terms are the $\vec{L}\Dot\vec{S}_2$ portion of Eq.\,(\ref{VLS}) (call it $V'_{LS}$) and $V_T$. Their matrix elements for a generic $2P$ state are
\begin{subequations}
\begin{eqnarray}
\langle 2P|V'_{LS}|2P\rangle &=& \frac{\mu^3\al^4(1+\ka)}{24m_1m_2} \left[1+\frac{m_1}{2m_2}\frac{(1+2\ka)}{(1+\ka)}\right]\langle\vec{L}\Dot\vec{S}_2\rangle \,,\\
\langle 2P|V'_T|2P\rangle &=& \frac{\mu^3\al^4(1+a_\mu)(1+\ka)}{24m_1m_2} \langle3\vec{S}_1\Dot\hat{r}\vec{S}_2\Dot\hat{r}-\vec{S}_1\Dot\vec{S}_2\rangle \,.
\end{eqnarray}
\end{subequations}

For $V'_{LS}$, we can use the fact that $\vec{F}^{\,2}$ and $\vec{J}^{\,2}$ are diagonal, so $\vec{F}^{\,2}=(\vec{J}+\vec{S}_2)^2$ and we can obtain the relation
\begin{equation} \label{LS2}
2\vec{L}\Dot\vec{S}_2=\vec{F}^{\,2}-\vec{J}^{\,2}-2\vec{S}_1\Dot\vec{S}_2-3/4 =\vec{F}^{\,2}-\vec{J}^{\,2}+3/4-\vec{S}^{\,2}\,.
\end{equation}
Both the $^5P_{3/2}$ and $^1P_{1/2}$ states are eigenstates of $\vec{S}^{\,2}$ with eigenvalue $2$, so $\langle ^5P_{3/2}|\vec{L}\Dot\vec{S}_2|^5P_{3/2}\rangle$ = 1/2 and \\ 
$\langle ^1P_{1/2}|\vec{L}\Dot\vec{S}_2|^1P_{1/2}\rangle$ = $-1$. 
\begin{subequations}
\begin{eqnarray}
\langle ^1P_{1/2}|V'_{LS}|^1P_{1/2}\rangle &=& -\frac{\mu^3\al^4(1+\ka)}{24m_1m_2}\left[1+\frac{1}{2} \frac{m_1}{m_2}\frac{(1+2\ka)}{(1+\ka)}\right] \\ [6pt]
\langle ^5P_{3/2}|V'_{LS}|^5P_{3/2}\rangle &=& \frac{\mu^3\al^4(1+\ka)}{48m_1m_2}\left[1+\frac{1}{2} \frac{m_1}{m_2}\frac{(1+2\ka)}{(1+\ka)}\right] 
\end{eqnarray}
\end{subequations}

The matrix elements of $V_T$ for these states are
\begin{subequations}
\begin{eqnarray}
\langle ^1P_{1/2}|V_T|^1P_{1/2}\rangle &=& -\frac{\mu^3\al^4(1+\ka)}{3m_1m_2}\frac{(1+a_\mu)}{8} \\ [6pt]
\langle ^5P_{3/2}|V_T|^5P_{3/2}\rangle &=& -\frac{\mu^3\al^4(1+\ka)}{3m_1m_2}\frac{(1+a_\mu)}{80}
\end{eqnarray}
\end{subequations}
Combining these two contributions gives
\begin{subequations}
\begin{eqnarray}
E_{HFS}(^1P_{1/2}) &=& -\frac{\mu^3\al^4(1+\ka)}{3m_1m_2}\left[\frac{1}{4}+\frac{a_\mu}{8}+\frac{1}{16} \frac{m_1}{m_2}\frac{(1+2\ka)}{(1+\ka)}\right]\to -5.968\, {\rm meV}\\[6pt]
E_{HFS}(^5P_{3/2}) &=& \frac{\mu^3\al^4(1+\ka)}{3m_1m_2}\left[\frac{1}{20}-\frac{a_\mu}{80}+\frac{1}{32} \frac{m_1}{m_2}\frac{(1+2\ka)}{(1+\ka)}\right]\to 1.272\, {\rm meV} 
\end{eqnarray}
\end{subequations}

The $^3P_{3/2}$ and $^3P_{1/2}$ states are are mixed by the $V'_{LS}$ and $V_T$ potentials. This results in
\begin{equation}
\langle ^3P_{1/2}|\vec{L}\Dot\vec{S}_2|^3P_{1/2}\rangle = 1/3\,, \quad \langle ^3P_{3/2}|\vec{L}\Dot\vec{S}_2|^3P_{3/2}\rangle = -5/6\,, \quad \langle ^3P_{3/2}|\vec{L}\Dot\vec{S}_2|^3P_{1/2}\rangle = -\sqrt{2}/3\,.
\end{equation}
and
\begin{eqnarray}
\langle ^3P_{1/2}|3\vec{S}_1\Dot\hat{r}\vec{S}_2\Dot\hat{r}-\vec{S}_1\Dot\vec{S}_2|^3P_{1/2}\rangle &=& 1/3\,,\quad \langle ^3P_{3/2}|3\vec{S}_1\Dot\hat{r}\vec{S}_2\Dot\hat{r}-\vec{S}_1\Dot\vec{S}_2|^3P_{3/2}\rangle = 1/6\,,\nonumber \\ [6pt]
\langle ^3P_{3/2}|3\vec{S}_1\Dot\hat{r}\vec{S}_2\Dot\hat{r}- \vec{S}_1\Dot\vec{S}_2|^3P_{1/2}\rangle &=& \sqrt{2}/6\,.
\end{eqnarray}
The matrix elements of $V'_{LS}$ are
\begin{subequations}
\begin{eqnarray}
\langle ^3P_{1/2}|V'_{LS}|^3P_{1/2}\rangle &=&\frac{\mu^3\al^4(1+\ka)}{72m_1m_2}\left[1+\frac{1}{2} \frac{m_1}{m_2}\frac{(1+2\ka)}{(1+\ka)}\right]\,, \\ [6pt]
\langle ^3P_{3/2}|V'_{LS}|^3P_{3/2}\rangle &=&  -\frac{5\mu^3\al^4(1+\ka)}{144m_1m_2}\left[1+\frac{1}{2} \frac{m_1}{m_2}\frac{(1+2\ka)}{(1+\ka)}\right]\,,\\ [6pt]
\langle ^3P_{3/2}|V'_{LS}|^3P_{1/2}\rangle &=&-\frac{\mu^3\al^4(1+\ka)}{72m_1m_2} \left[1+\frac{1}{2} \frac{m_1}{m_2}\frac{(1+2\ka)}{(1+\ka)}\right]\sqrt{2}\,.
\end{eqnarray}
\end{subequations}
and those of $V_T$ are
\begin{subequations}
\begin{eqnarray}
\langle ^3P_{1/2}|V_T|^3P_{1/2}\rangle &=& \frac{\mu^3\al^4}{72m_1m_2} (1+\ka)(1+a_\mu)\,, \\ 
\langle ^3P_{3/2}|V_T|^3P_{3/2}\rangle &=& \frac{\mu^3\al^4}{144m_1m_2} (1+\ka)(1+a_\mu)\,, \\ 
\langle ^3P_{3/2}|V_T|^3P_{1/2}\rangle &=& \frac{\mu^3\al^4}{144m_1m_2} (1+\ka)(1+a_\mu)\sqrt{2}\,.
\end{eqnarray}
\end{subequations}
The $V'_{LS}$ and $V_T$ contributions can be combined to give
\begin{subequations}
\begin{eqnarray}
E_{HFS}(^3P_{1/2}) &=& \frac{\mu^3\al^4(1+\ka)}{9m_1m_2}\left[\frac{1}{4}+\frac{a_\mu}{8}+\frac{1}{16} \frac{m_1}{m_2}\frac{(1+2\ka)}{(1+\ka)}\right]\to 1.989\, {\rm meV}\\[6pt]
E_{HFS}(^3P_{3/2}) &=& -\frac{5\mu^3\al^4(1+\ka)}{18m_1m_2}\left[\frac{1}{10}-\frac{a_\mu}{40}+\frac{1}{16} \frac{m_1}{m_2}\frac{(1+2\ka)}{(1+\ka)}\right]\to -2.120\, {\rm meV}\\[6pt]
E_{MIX}(^3P_{1/2}-^3P_{3/2}) &=& -\frac{\mu^3\al^4(1+\ka)}{3m_1m_2}\left[1+ \frac{m_1}{m_2}\frac{(1+2\ka)}{(1+\ka)}\right]\frac{\sqrt{2}}{48}\to -0.796\, {\rm meV}\label{MIX}
\end{eqnarray}
\end{subequations}
The expression for $E_{MIX}$ omits the $a_\mu$ contribution and all these results agree with Ref.\cite{Pachucki}.

Diagonalizing the triplet $P$ mixing matrix
\begin{equation}
\left(\begin{array}{cc}
E_{HFS}(^3P_{1/2}) & E_{MIX} \\ [6pt]
E_{MIX}    & E_{HFS}(^3P_{3/2})+\Delta E_{FS}
\end{array}\right)
\end{equation}
has the effect of shifting the $^3P_{3/2}$ level up by $\Delta = 0.1447\,{\rm meV}$ and the $^3P_{1/2}$ level down by the same amount.

There are small electron vacuum polarization corrections to all of the terms in the potential that contribute to order $\al^4$. These are computed in the Appendix and included in the results that are compared with experiment.

\section{Results and conclusions}
Relative to the $n=2$ Bohr level, the energies of the various $n=2$ states, including the small corrections calculated in the Appendix, are
\begin{subequations}
\begin{eqnarray}
E(^5P_{3/2}) &=& EE_{VP}+E_2+E_4+E_{FS}(P_{3/2})+E_{HFS}(^5P_{3/2})=215.609\, {\rm meV} \\
E(^3P_{3/2}) &=& EE_{VP}+E_2+E_4+E_{FS}(P_{3/2})+E_{HFS}(^3P_{3/2})+\Delta =212.360\,{\rm meV} \\
E(^3P_{1/2}) &=& EE_{VP}+E_2+E_4+E_{FS}(P_{1/2})+E_{HFS}(^3P_{1/2})-\Delta =207.826\,{\rm meV} \\
E(^1P_{1/2}) &=& EE_{VP}+E_2+E_4+E_{FS}(P_{1/2})+E_{HFS}(^1P_{1/2}) =200.006\,{\rm meV}\\
E(^3S_{1/2}) &=& E_{HF}(^3S_{1/2})=5.7351\,{\rm meV}\\
E(^1S_{1/2}) &=& E_{HF}(^1S_{1/2})=-17.2054\,{\rm meV}\\
E(P_{1/2})-E(S_{1/2}) &=& EE_{VP}+E_2+E_4+E_{FS}(P_{1/2})=205.980\,{\rm meV} \label{Lamb}
\end{eqnarray}
\end{subequations}
Here, $EE_{VP}$ is the sum of the first seven rows in Table I, $E_2$ is the eight row, $E_4$ the ninth row, $E_{FS}$ the tenth row, the s-state hyperfine splitting the eleventh row and the p-state hyperfine splittings the twelfth row. The additional small corrections in the Appendix include the spin independent terms in subsection 1, the $\ell=0$ hyperfine splitting in subsection 4, the fine structure splittings in subsection 2, and the $\ell=1$ hyperfine splitting in subsections 3,4 and 5. Eq.\,(\ref{Lamb}) gives our value for the Lamb shift in the absence of proton structure corrections.

Including all the contributions summarized in Tables I and II, the theoretical expression for the $2^3S_{1/2}\leftrightarrow 2^5P_{3/2}$ interval compared to the experimental result is
\begin{subequations}
\begin{eqnarray}
209.8740\,{\rm meV} - 5.2257\,r_p^2 \;{\rm meV\,fm^{-2}}(+0.035\,r_p^3\;{\rm meV\,fm^{-3}})&=& 206.2925\,{\rm meV} \\ [6pt]
r_p &=& 0.8276\;{\rm fm}\;(0.8302\;{\rm fm}) \label{comp1} \,,
\end{eqnarray}
\end{subequations}
where $r_p=\sqrt{\langle r^2\rangle}$ denotes the proton mean square radius and the terms in parenthesis include an estimate of the proton polarizability correction \cite{MF1,MF2,PH,GP}.
The corresponding result for the $2^1S_{1/2}\leftrightarrow 2^3P_{3/2}$ interval is
\begin{subequations}
\begin{eqnarray}
229.5652\,{\rm meV} - 5.2257 r_p^2\;{\rm meV\,fm^{-2}}(+0.035\,r_p^3\;{\rm meV\,fm^{-3}})&=& 225.8535\,{\rm meV}\\ [6pt]
r_p &=& 0.8428\;{\rm fm}\,\;(0.8452\;{\rm fm})\label{comp2}\,.
\end{eqnarray}
\end{subequations}
These results are consistent with each other but low compared to other theoretical calculations. Reference \cite{nature} makes a well reasoned argument that theoretical value for the $2^3S_{1/2}\leftrightarrow 2^5P_{3/2}$ transition is  $209.978\,{\rm meV}$ or about $0.1\,{\rm meV}$ greater than the present result. It gives the 0.841 fm result and one needs less than an additional 0.2 meV to agree with the electron scattering result whereas we need something closer to an additional 0.3 meV. 

\begin{center}
\begin{table}[h]
\begin{tabular}{|l|r|r|r|r|r|r|r|}
\hline
\st & \mc{1}{c|}{$^1S_{1/2}$}& \mc{1}{c|}{$^3S_{1/2}$}& \mc{1}{c|}{$^1P_{1/2}$}& \mc{1}{c|}{$^3P_{1/2}$}& \mc{1}{c|}{$^3P_{3/2}$}& \mc{1}{c|}{$^5P_{3/2}$}& \mc{1}{c|}{$^3P_{1/2}\leftrightarrow\, ^3P_{3/2}$} \\
\hline
\st one loop vacuum polarization & & &205.007 &205.007 &205.007 &205.007& \\
\hline
\st relativistic one loop correction  & & &0.021 &0.021 &0.026 &0.026 & \\
\hline
\st two-loop vacuum polarization & & &1.508 &1.508 &1.508 &1.508 & \\
\hline
\st relativistic two loop correction  & & &0.0001 &0.0001 &0.0002 &0.0002 & \\
\hline
\st three-loop vacuum polarization & & &0.0044 &0.0044 &0.0044 &0.0044 & \\
\hline 
\st NR 2nd order $\al^4$ vacuum polarization & & &0.1509 &0.1509 &0.1509 &0.1509 & \\ 
\hline
\st NR 2nd order $\al^5$ vacuum polarization & & & 0.0022 & 0.0022 & 0.0022 & 0.0022 & \\
\hline
\st spin independent $\al^4$ & & &5.6102 &5.6102 &5.6102 &5.6102 &\\
\hline
\st spin independent $\al^5$ & & &-0.7073 &-0.7073 &-0.7073 &-0.7073 &\\
\hline
\st fine structure & & &-5.5645 &-5.5645 &2.7823 &2.7823 & \\
\hline
\st hyperfine $\ell=0$ &-17.113 &5.7045 & & & & & \\
\hline
\st hyperfine $\ell=1$ & & &-5.9682 &1.9894 &-2.1195 &1.2717 & \\
\hline
\st mixing matrix element & & & & & & &-0.79615 \\
\hline
\mc{1}{|c|}{\st}&\mc{7}{c|}{Vacuum polarization corrections to $V_2(\vec{k})$} \\
\hline
\st & -0.0920 & 0.0307 &-0.0532 &-0.0464 &-0.0446 &-0.0418 &-0.0003 \\
\hline
\st total & -17.205 &5.735 &200.006 &207.971 &212.215 &215.609 &-0.7965 \\
\hline
\mc{4}{|l|}{\st result of mixing} &207.826 &212.360 &\mc{2}{c|}{} \\
\hline
\end{tabular}
\caption{The entries summarize the various corrections (in meV) to the $n=2$ states of muonic hydrogen calculated using Eqs.\,(\ref{VHF}-\ref{VSI}). }
\end{table}
\begin{table}[h]
\begin{tabular}{|l|r|}
\hline
\mc{2}{|c|}{\st $\langle r^2\rangle$ contributions} \\
\hline
\st leading order & $-5.1975\langle r^2 \rangle\;{\rm meV\,fm^{-2}}$ \\
\hline
\st first order propagator & $-0.0112\langle r^2 \rangle\;{\rm meV\,fm^{-2}}$ \\
\hline
\st second order correction & $-0.01695 \langle r^2\rangle\;{\rm meV\,fm^{-2}}$ \\
\hline
\st total & $-5.2257\langle r^2\rangle\;{\rm meV\,fm^{-2}}$ \\
\hline
\end{tabular}
\caption{Contributions related to the proton mean squared radius are listed. The corrections are given in equations (A61) and (A64). Proton polarizability corrections are not included.}
\end{table}
\end{center}
\vspace{-16pt}

We have used nonrelativistic wavefunctions throughout because our potential contains the leading order relativistic, recoil and one-loop corrections.  However, if we use the solutions to the Dirac equation given in Rose \cite{Rose}, the value of the electron one loop vacuum polarization changes from 205.007 meV to 205.028 meV for the $2P_{1/2}-2S_{1/2}$ interval and from 205.007 meV to 205.033 meV for the $2P_{3/2}-2S_{1/2}$ interval \cite{Borie_1}. We have also calculated the relativistic corrections to the two-loop vacuum polarization contribution using the approach of \cite{KN}. In this order, the change in the $2P_{1/2}-2S_{1/2}$ interval is 0.0001 meV and the corresponding change in the $2P_{3/2}-2S_{1/2}$ interval is 0.0002 meV.  

A more fundamental way to calculate the relativistic corrections to the dominant electron vacuum polarization contribution would be to use the solutions to the Salpeter equation with an instantaneous Coulomb kernel. Estimates of this correction using the scalar Salpeter equation have been made \cite{Jent} and the results are small. Unfortunately, analytic solutions for the spin $1/2$ Salpeter wave functions with unequal masses are not available.

The difference in the values of $r_p$ given in (\ref{comp1}) and (\ref{comp2}) suggest that a reasonable estimate of the error (deviation from the average) in $r_p$ would be $\pm 0.005$.  This leaves a substantial difference from the CODATA value of $r_p  = 0.8775(51)\;{\rm fm}$ obtained from electron-proton scattering \cite{MTN}.

Finally, one might wonder how the relatively large contributions from the mass dependence of the delta function term and the one loop $\al^2/r^2$ term mentioned in the Introduction can still lead to a Lamb shift value that is in agreement with other calculations. The answer is that there are two versions of the spin-independent fine structure Hamiltonian that contribute order $\al^4$ corrections to the $2P_{1/2}-2S_{1/2}$ energy difference. One is the Breit-Pauli version \cite{Pohl} $H'_{B-P}$ given by 
\begin{eqnarray}
H'_{B-P}&=&\frac{\pi\al}{2}\left(\frac{1}{m_1^2}+\frac{1}{m_2^2}\right) \de(\vec{r})-\frac{\al}{2m_1m_2}p_i\left(\frac{\de_{ij}}{r} +\frac{x_ix_j}{r^3}\right)p_j \label{BP1} \\
&=&\frac{\pi\al}{2\mu^2}\de(\vec{r})-\frac{\pi\al}{m_1m_2}\de(\vec{r}) -\frac{\al}{2m_1m_2}p_i\left(\frac{\de_{ij}}{r} +\frac{x_ix_j}{r^3}\right)p_j \label{BP2}\,.
\end{eqnarray}
The other is $H'_{GRS}$ of reference \cite{GRS}, which has the form
\begin{equation}
H'_{GRS}=\frac{\pi\al}{2\mu^2}\de(\vec{r})-\frac{\al}{m_1m_2}\frac{1}{r} \vec{p}^{\;2}+\frac{\mu\al^2}{2m_1m_2r^2}\label{GRS3}\,,
\end{equation}
where the last term arises in the calculation of the one-loop corrections. The last two terms of Eqs.\,(\ref{BP2}) and (\ref{GRS3}) give identical contributions to the $2P_{1/2}-2S_{1/2}$ splitting, namely
\begin{equation}
\Delta E(2P-2S)=\frac{\mu^3\al^4}{4m_1m_2}\,.
\end{equation}
All the spin-dependent fine structure terms of the two versions are the same so, apart from minor differences in some recoil terms, the Lamb shift values agree. This implies that there should be no $\al^2/r^2$ term associated with the one-loop corrections to Breit-Pauli Hamiltonian. In Appendix B this is shown to be the case.
\acknowledgments
W.~W.~R. would like to thank Levere Hostler, who introduced him to the properties of the Coulomb Green's function and Stanley Radford for a discussion about effective potentials.

\appendix
\section{Electron vacuum polarization corrections to order $\al^4$ terms}
The electron vacuum polarization corrections to the leading order $\al^4$ terms in the potential can be obtained from the momentum space representation of the potential, which is 
\begin{eqnarray}
V_2(\vec{k})&=&-\frac{e^2}{\vec{k}^{\,2}}\left[1-\frac{\vec{k}^{\,2}}{8\mu^2}+\frac{i}{4m_1m_2}\left (\left(2+\frac{m_2}{m_1}\right) (\vec{k}\times\vec{p})\Dot\vec{\sig}_1+(1+\ka)\left(2+\frac{1+2\ka}{1+\ka}\frac{m_1}{m_2}\right) (\vec{k}\times\vec{p}) \Dot\vec{\sig}_2\right)\right. \nonumber \\ 
&&\left.-\frac{(1+\ka)}{6m_1m_2}\vec{k}^{\,2}\vec{\sig}_1\Dot\vec{\sig}_2+\frac{(1+\ka)}{4m_1m_2} \left(\vec{k}\Dot\vec{\sig}_1\vec{k}\Dot\vec{\sig}_2-\frac{1}{3}\vec{k}^{\,2}\vec{\sig}_1 \Dot\vec{\sig}_2\right)+\frac{\vec{p}^{\;2}}{m_1m_2}-\frac{\al\mu\pi k}{4m_1m_2}-\frac{1}{6}\langle r^2\rangle\vec{k}^{\,2}\right]\,,
\end{eqnarray}
where the next to last term is a one-loop correction that contributes at order $\al^4$. The electron vacuum polarization correction is obtained by making the replacement
\begin{equation}
-\frac{e^2}{\vec{k}^{\,2}}\rightarrow -e^2\int_{4m_e^2}^\infty\!\!\frac{d\la}{\la}\frac{\Delta^{(2)}(\la)}{\vec{k}^{\,2}+\la}\,.
\end{equation}
In addition, there is a second order perturbative correction containing $V_{VP}(r)$ for each of these terms as well as for the relativistic kinetic energy terms. In these calculations, all integrals except those over $\la$ can be evaluated analytically using Mathematica. The integrals over $\la$ are performed using the Mathematica {\tt NIntegrate} routine.  
\subsection{Spin independent terms}
\noindent\fbox{$e^2\vec{k}^{\,2}/8\mu^2$} \\[6pt]
This leads to the expression
\begin{equation}
\frac{\pi\al}{2\mu^2}\int_{4m_e^2}^\infty\!\!\frac{d\la\,\Pi_e(\la)}{\la}\frac{\vec{k}^{\,2}}{\vec{k}^{\,2}+\la}\,,
\end{equation}
and transforming to coordinate space gives
\begin{equation}
\frac{\pi\al}{2\mu^2}\int_{4m_e^2}^\infty\!\!\frac{d\la\,\Pi_e(\la)}{\la}\left(\de(\vec{r})-\frac{\la}{4\pi} \frac{1}{r}e^{-\sqrt{\la}r}\right).
\end{equation}
The delta function only contributes to $\ell=0$ and gives
\begin{equation}
\Delta E(\de)=\frac{\mu\al^5}{48\pi}\int_4^\infty\!\! \frac{dx}{x}(1+2/x)\sqrt{1-4/x}\,.
\end{equation}
The integral diverges, but the $\ell=0$ contribution of the remaining term cancels this divergence. Using Eq.\,(14) and the extra factor of $\la=m_e^2x$, the remaining contribution to the $E(2P)-E(2S)$ interval is
\begin{equation}
\Delta (E(2P)-E(2S))=\frac{\mu\al^5}{48\pi}\int_4^\infty\!\! \frac{dx\,m_e^4a^4x^2(1+2/x)\sqrt{1-4/x}}{x(1+m_ea\sqrt{x})^4}\,.
\end{equation}
The vacuum polarization correction to this term is
\begin{equation}\label{delta corr}
\Delta E_1(e^2\vec{k}^{\,2}/8\mu^2)=\frac{\mu\al^5}{48\pi}\int_4^\infty\!\! \frac{dx}{x}(1+2/x)\sqrt{1-4/x}\left[\frac{m_e^4a^4x^2}{(1+m_ea\sqrt{x})^4} -1\right]=-0.0363\,{\rm meV}\,.
\end{equation}
The second order correction is
\begin{eqnarray}
\Delta E_2(e^2\de(\vec{r})/8\mu^2) &=& 2\left(\frac{\pi\al}{2\mu^2}\right)\int\!\!d^3r\, \psi_{20}(\vec{r})V_{VP}(r)\int\!\!dr'^3 g_{20}(r,r')\de(\vec{r})\psi_{20}(\vec{r}) \nonumber \\
&=&-\left(\frac{\pi\al^2}{\mu^2}\right)\int\!\!d^3r\,\frac{R_{20}(r)}{8\pi a^3}\frac{1}{r}e^{-\sqrt{\la}r}g_{20}(r/a,0) \nonumber \\
&=& -\left(\frac{\pi\al^2}{2\mu^2\,a^3}\right)\int_0^\infty\!\!dr\, r(1-\frac{r}{2a})e^{-r/2a}e^{-\sqrt{\la}\,r}g_{20}(r/a,0) \nonumber \\
\Delta E_2(e^2\de(\vec{r})/8\mu^2) &=& -\left(\frac{\pi\al^3}{2\mu}\right) \int_0^\infty\!\!dx\,x(1-x/2)e^{-x/2}e^{-\sqrt{\la}\,a\,x}g_{20}(x,0)\,,
\end{eqnarray}
where the integration over $d\la$ has been suppressed. The Greens function $g_{20}(x,0)$ is
\begin{equation}
g_{20}(x,0)=\frac{\mu^2\al}{4\pi}\frac{e^{-x/2}}{2x}\left[-4+(8\ga-6)x+(13-4\ga)x^2-x^3 -4x(x-2)\ln(x)\right]=\frac{\mu^2\al}{4\pi}{\cal G}_{20}(x)\,.
\end{equation}
The final expression for the $2p-2s$ splitting is 
\begin{equation}
-\Delta E_2(e^2\de(\vec{r})/8\mu^2)=\frac{\mu\al^4}{8}\frac{\al}{3\pi} \int_4^\infty\!\!\frac{dx}{x}(1+2/x)\sqrt{1-4/x}\,F(\be\sqrt{x})=-0.0549\;{\rm meV}\,,
\end{equation}
where
\begin{eqnarray}
F(\be\sqrt{x})&=&\int_0^\infty\!\!dy\,y(1-y/2)e^{-y/2} e^{-\be\sqrt{x}\,y}{\cal G}_{20}(y) \nonumber \\
F(\be\sqrt{x})&=&-\frac{3+11\be\sqrt{x}+4\be^2x+12\be^3x^{3/2}+4\be^4x^2 +4(1+\be\sqrt{x})(1+2\be^2x)\ln(1+\be\sqrt{x})}{2(1+\be\sqrt{x})^5}\label{intg}\,,
\end{eqnarray}
and $\be=m_ea$.

\noindent\fbox{$-e^2\vec{p}^{\;2}/rm_1m_2$} \\ [6pt]
Here, the expression is
\begin{equation}
\frac{-e^2}{m_1m_2}\frac{1}{(2\pi)^3}\int\!\!\frac{d^{\,3}k\,e^{i\vec{k}\cdot\vec{r}}} {\vec{k}^{\,2}+\la}\vec{p}^{\;2}=-\frac{\al}{m_1m_2r}e^{-\sqrt{\la}r}\vec{p}^{\;2} =-\frac{2\mu\al^2}{m_1m_2r^2} \left(1-\frac{r}{8a} \right)e^{-\sqrt{\la}r}\,\,.
\end{equation}
For the $\ell=1$ state, the result of integrating over $r$ is
\begin{equation}
\frac{\mu^3\al^4}{12m_1m_2}\frac{1}{(1+a\sqrt{\la})^4}\left(\frac{3}{4}-2(1+a\sqrt{\la})\right)\,.
\end{equation}
The $\ell=0$ state, when integrated over $r$ gives
\begin{equation}
\frac{\mu^3\al^4}{m_1m_2}\frac{1}{(1+a\sqrt{\la})^4}\left(-(1+a\sqrt{\la})^3+\frac{9}{8}(1+a\sqrt{\la})^2 -\frac{3}{4}(1+a\sqrt{\la})+\frac{3}{16}\right)\,.
\end{equation}
Integrating the difference of these two results over $\la$ gives
\begin{equation}
\Delta E_1(-e^2\vec{p}^{\;2}/rm_1m_2)=0.0142\,{\rm meV}\,.
\end{equation}
The expression for the second order correction is (again, suppressing the integral over $d\la$ and including the factor of 2)
\begin{equation}
\Delta E_2(-e^2\vec{p}^{\;2}/rm_1m_2)= \frac{4\mu^3\al^4}{m_1m_2}\int_0^\infty\!\!drr R_{2\ell}(r)e^{-\sqrt{\la}\,r}\int_0^\infty\!\!dr'g_{2\ell}(r/a,r'/a)(1-r'/8a)R_{2\ell}(r')\,. \end{equation}
This expression reduces to
\begin{equation}
\frac{\mu^3\al^4}{6m_1m_2}\int_0^\infty\!\!dxx^2e^{-x/2}e^{-\sqrt{\la}\,a\,x}\int_0^\infty \!\!dx'g_{21}(x,x')x'e^{-x'/2}(1-x'/8)\,,
\end{equation}
for $\ell=1$ and
\begin{equation}
\frac{2\mu^3\al^4}{m_1m_2}\int_0^\infty\!\!dxx(1-x/2)e^{-x/2}e^{-\sqrt{\la}\,a\,x} \int_0^\infty\!\!dx'g_{20}(x,x')(1-x'/2)e^{-x'/2}(1-x'/8)\,,
\end{equation}
for $\ell=0$.
The integrals are, respectively
\begin{equation}
-\frac{13+33a\sqrt{\la}+32(1+a\sqrt{\la})\ln(1+a\sqrt{\la})}{8(1+a\sqrt{\la})^5}\,,
\end{equation}
and
\begin{equation}
-\frac{30+86a\sqrt{\la}-60a^2\la+52a^3\la^{3/2}+64(1+a\sqrt{\la})(1+2a^2\la) \ln(1+a\sqrt{\la})}{64(1+a\sqrt{\la})^5}\,.
\end{equation}
Integrating the difference over $d\la$ gives a $2p-2s$ contribution of
\begin{equation}
\Delta E_2(-e^2\vec{p}^{\;2}/rm_1m_2)=0.01459\;{\rm meV}.
\end{equation}
\fbox{$-(\vec{p}^{\,2})^2/8m_1^3$} \\ [6pt]
When integrated over $d\la$, the second order correction to the relativistic kinetic energy contribution is
\begin{equation}
\Delta E_2(-(\vec{p}^{\,2})^2/8m_1^3)=\frac{4\mu^4\al^4}{8m_1^3}\int_0^\infty \!\!dr\,r\,R_{2\ell}(r)e^{-\sqrt{\la}\,r}\int_0^\infty\!\!dr'\,g_{2\ell}(r/a,r'/a) (1-r'/8a)^2R_{2\ell}(r')\,.
\end{equation}
This correction is negligible for $\ell=1$. After completion of the integrals above, the contribution from the muon to the $2p-2s$ splitting for $\ell=0$ is
\begin{equation}
\Delta E_2(-(\vec{p}^{\,2})^2/8m_1^3) =\frac{\mu^4\al^4}{2m_1^3} \frac{\al}{3\pi}\int_4^\infty\!\!\frac{dx}{x}(1+2/x)\sqrt{1-4/x}\,G(\be\sqrt{x}) =0.02237\;{\rm meV}\,,
\end{equation}
with
\begin{equation}
G(\be\sqrt{x})=-\frac{7+\be\sqrt{x}(19+2\be\sqrt{x}(-7+5\be\sqrt{x}))+16(1+\be\sqrt{x}) (1+2\be^2x)\ln(1+\be\sqrt{x})}{16(1+\be\sqrt{x})^5}\,.
\end{equation}\\[6pt]
\fbox{$\al\mu\pi k/4m_1m_2$} \\ [6pt]
Here, the momentum space integral is
\begin{equation}
\frac{\mu\al^2\pi^2}{m_1m_2}\frac{1}{(2\pi)^3}\int\!\!\frac{d^{3}kk\,e^{i\vec{k}\cdot\vec{r}} e^{-\ep k}}{\vec{k}^{\,2}+\la}=\frac{\mu\al^2}{2m_1m_2r^2} \left(1-\sqrt{\la}r\int_0^\infty\!\!\frac{dt}{t^2+1}\sin(\sqrt{\la}\,t\,r)\right)\,,
\end{equation}
where $\ep$ is taken to 0 after the integral is evaluated. The $dt$ integral can be evaluated at this point, but it is more convenient to first integrate over $dr$ with the integrand multiplied by $R_{2\ell}^2(r)$. For the $\ell =1$, the result is 
\begin{equation}
\frac{1}{24a^3}\int_0^\infty\!\!dr\left(1-\sqrt{\la}\int_0^\infty\!\!\frac{dt}{t^2+1}r\sin(\sqrt{\la}t\,r)\right) \frac{r^2}{a^2}e^{-r/a}=\frac{1}{24a^3}\left[2+\int_0^\infty\!\!\frac{dt}{t^2+1} \frac{24\la a^2t(\la a^2\,t^2-1)}{(1+\la a^2\,t^2)^4}\right]\,.
\end{equation}
The integration over $dt$ then gives
\begin{equation}
\frac{1}{12a^2}\left[1-\frac{\la a^2(\la^3a^6-3\la^2a^4+15\la a^2-6(\la a^2+1)\ln(\la a^2)-13)}{(1-\la a^2)^4}\right]\,.
\end{equation}
The calculation for $\ell=0$ is similar. Taking the difference and integrating over $\la$ results in the expression 
\begin{eqnarray}
\Delta E_1(e^2\al\mu\pi\,k/4m_1m_2) &=& \frac{\mu^3\al^5}{72\pi m_1m_2}\int_4^\infty\!\!\frac{dx}{x}\frac{(1+2/x)\sqrt{1-4/x}}{(1-\be^2x)^4}\left[26\be^6x^3-30\be^4x^2+6\be^2x\right. \nonumber \\ [6pt]
&&\left.-24\be^4x^2(\be^2x+1)\ln(\be\sqrt{x})-2\right]=-0.0030\,{\rm meV}\,,
\end{eqnarray}
where $\be=m_ea$.

The second order correction for $\mu\al^2/2m_1m_2r^2$ is obtained by integrating
\begin{equation}
\Delta E_2(\ell)=-\frac{\mu^3\al^4a^3}{2m_1 m_2}\int_0^\infty \!\!dx\,x\,R_{2\ell}(x)e^{-\sqrt{\la}\,a\,x}\int_0^\infty\!\! dx'\,g_{2\ell}(x,x')R_{2\ell}(x')\,,
\end{equation}
over $\la$. The results are
\begin{equation}
\Delta E_2(P)=0.00003\;{\rm meV}, \qquad \Delta E_2(S)= 0.00154\;{\rm meV}\,,
\end{equation}
for a $2p-2s$ splitting contribution of
\begin{equation}
\Delta E_2(e^2\al\mu\pi\,k/4m_1m_2) = -0.0015 \;{\rm meV}\,.
\end{equation} \\ [6pt]
\subsection{Fine structure}
\fbox{$-e^2ik_i/4m_1m_2$} \\ [6pt]
The integral to be evaluated is
\begin{equation}
\frac{1}{(2\pi)^3}\int\!\frac{dk^{3}\,k_i\,e^{i\vec{k}\cdot\vec{r}}}{\vec{k}^{\,2}+\la}=i\frac{x_i}{4\pi r^3}\left(1+\sqrt{\la}\,r\right)e^{-\sqrt{\la}\,r}\,.
\end{equation}
This leads to the spin-orbit contributions
\begin{equation}
\frac{\al}{2m_1m_2\,r^3}\left[\left(2+\frac{m_2}{m_1}\right)\vec{L}\Dot\vec{S}_1+ (1+\ka)\left(2+\frac{(1+2\ka)m_1}{(1+\ka)m_2}\right)\vec{L}\Dot\vec{S}_2\right] \left(1+\sqrt{\la}\,r\right)e^{-\sqrt{\la}\,r}\,.
\end{equation}
Only the $p$ state is affected and we have
\begin{equation}
\langle 2p|\frac{1}{r^3}\left(1+\sqrt{\la}\,r\right)e^{-\sqrt{\la}\,r} |2p\rangle=\frac{1}{24a^3}\frac{1+3a\sqrt{\la}}{(1+a\sqrt{\la})^3}\,.
\end{equation}
For the $\vec{L}\Dot\vec{S}_1$ term, $\langle \vec{L}\Dot\vec{S}_1\rangle$ is $1/2$ for the $P_{3/2}$ states and $-1$ for the $P_{1/2}$ states. Thus, their (fine structure) contributions are
\begin{eqnarray} \label{DeltaE_FS}
&&\frac{\mu^3\al^5}{144\pi m_1^2} (1+2m_1/m_2)\langle\vec{L}\Dot\vec{S}_1\rangle\int_4^\infty\!\!\frac{dx}{x}\,\frac{(1+2/x)\sqrt{1-4/x}(1+3\be\sqrt{x})} {(1+\be\sqrt{x})^3}\nonumber \\
&&=\left\{\begin{array}{l}
                     \Delta E_{1\;FS}(P_{3/2})=0.001\;{\rm meV} \\ [6pt]
                     \Delta E_{1\;FS}(P_{1/2})=-0.002\;{\rm meV}
                     \end{array}\right.\,.
\end{eqnarray}
The second order contribution can be obtained from the expression
\begin{equation}
\Delta E_{2\;FS}(P)= -\frac{\mu^3\al^4}{24m_1^2}(1+2m_1/m_2)\langle\vec{L}\Dot\vec{S}_1\rangle \int_0^\infty\!\!dt\,t^2e^{-t/2}e^{-\sqrt{\la}\,a\,t}\int_0^\infty\!\!dt'g_{21}(t,t') e^{-t'/2}\,,
\end{equation}
integrated over $\la$. The result is
\begin{eqnarray}\label{DeltaE2 FS}
&&-\frac{\mu^3\al^5}{72m_1^2}(1+2m_1/m_2) \langle\vec{L}\Dot\vec{S}_1\rangle\int_4^\infty\!\!\frac{dx}{x}\,(1+2/x)\sqrt{1-4/x}\,H(\be) \nonumber \\
&&=\left\{\begin{array}{l}
                     \Delta E_{2\;FS}(P_{3/2})=0.0006\;{\rm meV} \\ [6pt]
                     \Delta E_{2\;FS}(P_{1/2})=-0.0012\;{\rm meV}
                     \end{array}\right.\,.
\end{eqnarray}
with
\begin{equation}\label{Hbeta}
H(\be\sqrt{x})=-\frac{3+\be\sqrt{x}(11+4\be\sqrt{x})+4(1+\be\sqrt{x})\ln(1+\be\sqrt{x})} {2(1+\be\sqrt{x})^5}\,.
\end{equation}
\subsection{p state hyperfine splitting}

The $\vec{L}\Dot\vec{S}_2$ term gives corrections to the hyperfine splitting. The values of these corrections are obtained from Eq.\,(\ref{DeltaE_FS}) by changing the coefficient of the integral to
\begin{equation}
\frac{\mu^3\al^5}{144\pi m_1m_2}(1+\ka) \left(2+\frac{(1+2\ka)}{(1+\ka)}\frac{m_1}{m_2}\right) \langle\vec{L}\Dot\vec{S}_2\rangle
\end{equation}
and using
\begin{subequations}
\begin{eqnarray}
\langle ^5P_{3/2}|\vec{L}\Dot\vec{S}_2|^5P_{3/2}\rangle &=&\f12\,, \quad \langle ^3P_{3/2}|\vec{L}\Dot\vec{S}_2|^3P_{3/2}\rangle = -\f56 \\ [6pt]
\langle ^3P_{1/2}|\vec{L}\Dot\vec{S}_2|^3P_{1/2}\rangle &=&\f13\,, \quad
\langle ^1P_{1/2}|\vec{L}\Dot\vec{S}_2|^1P_{1/2}\rangle = -1 \\ [6pt]
\langle ^3P_{3/2}|\vec{L}\Dot\vec{S}_2|^3P_{1/2}\rangle &=& -\sqrt{2}/3\,.
\end{eqnarray}
\end{subequations}
The corresponding energy corrections are
\begin{subequations}
\begin{eqnarray}
\Delta E_{1\;HF}(^5P_{3/2}) &=& 0.0006\, {\rm meV}\,, \quad \Delta E_{1\;HF}(^3P_{3/2}) = -0.0009\, {\rm meV} \\ [6pt]
\Delta E_{1\;HF}(^3P_{1/2}) &=& 0.0004\, {\rm meV}\,, \quad \Delta E_{1\;HF}(^1P_{1/2}) = -0.0011\, {\rm meV}\\ [6pt]
\Delta E_{1\;HF}(Mix) &=&-0.0005\,{\rm meV}\,.
\end{eqnarray}
\end{subequations}

The second order perturbative hyperfine corrections can be obtained from Eq.\,(\ref{DeltaE2 FS}) by replacing its coefficient with
\begin{equation}
-\frac{\mu^3\al^5(1+\ka)}{72m_1m_2}\left(2+\frac{(1+2\ka)}{(1+\ka)}\right) \langle\vec{L}\Dot\vec{S}_2\rangle\,.
\end{equation}
The results are
\begin{subequations}
\begin{eqnarray}
\Delta E_{2\;HF}(^5P_{3/2}) &=& 0.0004\, {\rm meV}\,, \quad \Delta E_{2\;HF}(^3P_{3/2}) = -0.0006\, {\rm meV} \\ [6pt]
\Delta E_{2\;HF}(^3P_{1/2}) &=& 0.0002\, {\rm meV}\,, \quad \Delta E_{2\;HF}(^1P_{1/2}) = -0.0007\, {\rm meV}\\ [6pt]
\Delta E_{2\;HF}(Mix) &=&-0.0003\,{\rm meV}\,.
\end{eqnarray}
\end{subequations}
\subsection{Hyperfine splitting}
\fbox{$e^2(1+\ka)\vec{k}^{\,2}\sig_1\Dot\sig_2/6m_1m_2$} \\ [6pt]
This term is very similar to the spin independent $\vec{k}^{\,2}$ contribution treated above. The spin dependence means that the $s$ states and $p$ states must be treated separately. For the $s$ state, we have
\begin{equation}
\Delta VS_{HF}(r)=\frac{\al(1+\ka)}{6m_1m_2}\int_{4m_e^2}^\infty\!\!\frac{d\la}{\la}\Pi_e(\la) \left[4\pi\de(\vec{r})-\frac{\la}{r}e^{-\sqrt{\la}\,r}\right]\sig_1\Dot\sig_2\,.
\end{equation}
The delta function gives
\begin{equation}
\frac{\mu^3\al^4(1+\ka)}{12m_1m_2}\sig_1\Dot\sig_2\int_{4m_e^2}^\infty\!\! \frac{d\la}{\la}\Pi_e(\la)\,,
\end{equation}
and the remaining term gives
\begin{equation}
-\frac{\mu^3\al^4(1+\ka)}{12m_1m_2}\sig_1\Dot\sig_2\int_{4m_e^2}^\infty\!\! \frac{d\la}{\la}\Pi_e(\la)\left[\frac{a^4\la^2+a^2\la/2}{(1+a\sqrt{\la})^4}\right]\,.
\end{equation}
The energy corrections are
\begin{eqnarray}
\Delta E_{1\;HF}(S) &=& \frac{\mu^3\al^5(1+\ka)}{9\pi\,m_1m_2}\langle \vec{S}_1\Dot \vec{S}_2\rangle \int_4^\infty\!\!\frac{dx}{x}\frac{(1+2/x)\sqrt{1-4/x}(1+4\be\sqrt{x} +11\be^2\,x/2+4\be^3x^{3/2})}{(1+\be\sqrt{x})^4}\label{k^2 integral}\nonumber \\ [6pt]
&=& \left\{\begin{array}{l}
          \Delta E_{1\;HF}(^3S_{1/2})=0.0121\,{\rm meV} \\ [6pt]
          \Delta E_{1\;HF}(^1S_{1/2})=-0.0362\,{\rm meV}
          \end{array}\right.\,.
\end{eqnarray}
The second order contribution is given by
\begin{eqnarray}
E_{2\;HF}(S)&=&-\frac{2\mu^3\al^4(1+\ka)}{3m_1m_2}\langle \vec{S}_1\Dot \vec{S}_2\rangle\,\frac{\al}{3\pi}\int_4^\infty\!\!\frac{dx}{x} (1+2x)\sqrt{1-4/x}F(\be\sqrt{x}) \nonumber\\ [6pt]
&=& \left\{\begin{array}{l}
          \Delta E_{2\;HF}(^3S_{1/2})=0.01859\,{\rm meV} \\ [6pt]
          \Delta E_{2\;HF}(^1S_{1/2})=-0.05579\,{\rm meV}
          \end{array}\right.\,.
\end{eqnarray}
where $F(\be\sqrt{x})$ is given by Eq.\,(\ref{intg}). (See \cite{Borie_1}.)

For the $p$ state,
\begin{equation}
\Delta E_{HF}(P) = -\frac{\mu^3\al^4(1+\ka)}{6m_1m_2}\int_4^\infty\!\frac{dx}{x}\frac{\Pi_e(x)\be^2x}{(1+\be\sqrt{x})^4}\, \langle\vec{S}_1\Dot \vec{S}_2\rangle\,
\end{equation}
Using the expectation values
\begin{subequations}
\begin{eqnarray}
\langle ^5P_{3/2}|\vec{S}_1\Dot \vec{S}_2|^5P_{3/2}\rangle &=&\f14\,, \quad \langle ^3P_{3/2}|\vec{S}_1\Dot \vec{S}_2|^3P_{3/2}\rangle = -\ff512 \\ [6pt]
\langle ^3P_{1/2}|\vec{S}_1\Dot \vec{S}_2|^3P_{1/2}\rangle &=&-\ff112\,, \quad
\langle ^1P_{1/2}|\vec{S}_1\Dot \vec{S}_2|^1P_{1/2}\rangle = \f14\\
\end{eqnarray}
\end{subequations}
the corresponding energy corrections are
\begin{subequations}
\begin{eqnarray}
\Delta E_{HFP}(^5P_{3/2}) &=& 0.0002\, {\rm meV}\,, \quad \Delta E_{HFP}(^3P_{3/2}) = -0.0004\, {\rm meV} \\ [6pt]
\Delta E_{HFP}(^3P_{1/2}) &=& -0.0001\, {\rm meV} \quad \Delta E_{HFP}(^1P_{1/2}) = 0.0002\, {\rm meV}\,.
\end{eqnarray}
\end{subequations}
\subsection{Tensor splitting}
\fbox{$e^2(1+\ka)(k_ik_j-\de_{ij}\vec{k}^{\,2}/3)/4m_1m_2$} \\ [6pt]
The relevant integral in this case is 
\begin{equation}
\frac{1}{(2\pi)^3}\int\!\frac{dk^3\,(k_ik_j-\vec{k}^{\,2}\de_{ij}/3)\,e^{i\vec{k}\cdot\vec{r}}} {\vec{k}^{\,2}+\la}=\frac{1}{4\pi\,r^3}(\de_{ij}-3\hat{x}_i\hat{x}_j)\left(1+\sqrt{\la}\,r+\frac{\la\,r^2}{3}\right) e^{-\sqrt{\la}r}\,.
\end{equation}
The expression to be integrated over $\la$ is
\begin{equation}
\langle 2P|\Delta V_T(\vec{r})|2P\rangle=\frac{\al(1+\ka)}{m_1m_2} \langle\frac{1}{r^3}\left(1+\sqrt{\la}\,r+\frac{\la\,r^2}{3}\right) e^{-\sqrt{\la}r}\rangle\langle 3\vec{S}_1\Dot\hat{r}\vec{S}_2\Dot\hat{r} -\vec{S}_1\Dot\vec{S}_2\rangle\,.
\end{equation}
Now,
\begin{equation}
\langle\frac{1}{r^3}\left(1+\sqrt{\la}\,r+\frac{\la\,r^2}{3}\right) e^{-\sqrt{\la}r}\rangle=\frac{1}{24a^3}\frac{(1+4a\sqrt{\la}+5a^2\la)}{(1+a\sqrt{\la})^4}\,,
\end{equation}
so
\begin{equation}
\langle 2P|\Delta V_T|2P\rangle=\frac{\mu^3\al^5(1+\ka)}{72\pi m_1m_2}\int_4^\infty\!\frac{dx}{x}\frac{(1+2/x)\sqrt{1-4/x}(1+4\be\sqrt{x}+5\be^2x)} {(1+\be\sqrt{x})^4}\langle 3\vec{S}_1\Dot\hat{r}\vec{S}_2\Dot\hat{r} -\vec{S}_1\Dot\vec{S}_2\rangle\,.
\end{equation}
Using,
\begin{subequations}
\begin{eqnarray}
\langle ^5P_{3/2}|3\vec{S}_1\Dot\hat{r}\vec{S}_2\Dot\hat{r} -\vec{S}_1\Dot\vec{S}_2|^5P_{3/2}\rangle &=&-\f15\,, \quad \langle ^3P_{3/2}|3\vec{S}_1\Dot\hat{r}\vec{S}_2\Dot\hat{r} -\vec{S}_1\Dot\vec{S}_2|^3P_{3/2}\rangle = \f16 \\ [6pt]
\langle ^3P_{1/2}|3\vec{S}_1\Dot\hat{r}\vec{S}_2\Dot\hat{r} -\vec{S}_1\Dot\vec{S}_2|^3P_{1/2}\rangle &=&\f13\,, \quad
\langle ^1P_{1/2}|3\vec{S}_1\Dot\hat{r}\vec{S}_2\Dot\hat{r} -\vec{S}_1\Dot\vec{S}_2|^1P_{1/2}\rangle = -2 \\ [6pt]
\langle ^3P_{3/2}|3\vec{S}_1\Dot\hat{r}\vec{S}_2\Dot\hat{r} -\vec{S}_1\Dot\vec{S}_2|^3P_{1/2}\rangle &=& \sqrt{2}/6\,,
\end{eqnarray}
\end{subequations}
the corrections are
\begin{subequations}
\begin{eqnarray}
\Delta E_{1\;T}(^5P_{3/2}) &=& -0.0003\, {\rm meV}\,, \quad \Delta E_{1\;T}(^3P_{3/2}) = 0.0003\, {\rm meV} \\ [6pt]
\Delta E_{1\;T}(^3P_{1/2}) &=& 0.0005\, {\rm meV}\,, \quad \Delta E_{1\;T}(^1P_{1/2}) = -0.0030\, {\rm meV}\\ [6pt]
\Delta E_{1\;T}(Mix) &=& 0.00035\,{\rm meV}\,.
\end{eqnarray}
\end{subequations}

The perturbative second order corrections are obtained using Eq.\,(\ref{Hbeta}) and
\begin{equation}
\Delta E_T = -\frac{\mu^3\al^5(1+\ka)}{36\pi m_1m_2} \int_4^\infty\!\!\frac{dx}{x} (1+2/x)\sqrt{1-4/x}\,H(\be\sqrt{x}) \langle 3\vec{S}_1\Dot\hat{r}\vec{S}_2\Dot\hat{r} -\vec{S}_1\Dot\vec{S}_2\rangle\,.
\end{equation}
The results are
\begin{subequations}
\begin{eqnarray}
\Delta E_{2\;T}(^5P_{3/2}) &=& -0.0001\, {\rm meV}\,, \quad \Delta E_{2\;T}(^3P_{3/2}) = 0.0001\, {\rm meV} \\ [6pt]
\Delta E_{2\;T}(^3P_{1/2}) &=& 0.0002\, {\rm meV}\,, \quad \Delta E_{2\;T}(^1P_{1/2}) = -0.0013\, {\rm meV}\\ [6pt]
\Delta E_{1\;T}(Mix) &=& 0.00015\,{\rm meV}\,.
\end{eqnarray}
\end{subequations}
\subsection{Proton size corrections}
\fbox{$e^2\vec{k}^2\langle r^2\rangle/6$} \\ [6pt]
The contribution from the proton mean square radius can be obtained from Eq.\,(\ref{delta corr}) by replacing the coefficient of the integral by
\begin{equation}
\frac{\mu^3\al^5 \langle r^2 \rangle}{36\pi}\,.
\end{equation}
This results in the correction
\begin{equation}
\Delta E_1(\langle r^2 \rangle)=-0.0112\langle r^2 \rangle\;{\rm meV\,fm^{-2}}\,.
\end{equation}

There is an additional contribution containing $\langle r^2\rangle$ from second order perturbation theory. It has the form
\begin{eqnarray}
\Delta E_2(\langle r^2\rangle)&=&2\mu^2\al\left(\frac{2\pi\al \langle r^2\rangle}{3}\right)\int\!\!d^3r\,\psi_{20}(\vec{r})V_{VP}(r)\int\!\!d^3r' \frac{g_{20}(r,r')}{4\pi}\,\de(\vec{r})\psi_{20}(\vec{r})\nonumber \\
&=& -\frac{\mu^3\al^3\langle r^2\rangle}{6}\int_0^\infty\!\!dx\,x(1-x/2)e^{-\sqrt{\la}x}e^{-x/2}g_{20}(x,0)\,,
\end{eqnarray}
where we have suppressed the integration over $\la$. Using
\begin{equation}
g_{20}(x,0)=\frac{1}{2x}e^{-x/2}\left[-x^3+ (13-4\ga)x^2 +(8\ga-6)x-4-4x(x-2)\ln(x)\right]\,,
\end{equation}
and integrating over $\la$ gives
\begin{equation}
\Delta E_2(\langle r^2\rangle)=-0.0169543 \langle r^2\rangle\;{\rm meV\,fm^{-2}}\,,
\end{equation}
giving a total correction of $-5.2257\langle r^2 \rangle\;{\rm meV\,fm^{-2}}$, in agreement with \cite{Pachucki}.
\section{One-loop effective potential}
To obtain the full one-loop effective potential $V_4$, one has to evaluate the one-loop corrections to the single photon exchange potential $V_2$ and calculate
\begin{equation} \label{oneloop}
V_4(\vec{r})=\frac{1}{(2\pi)^3}\int\,d^{\,3}k\,e^{i\vec{k}\cdot\vec{r}} \left(\sum_{i}V_{4\,i}(\vec{k})\,-\,\de(V_2,V_2)\right)\,,
\end{equation}
where $i$ is the sum over all one-loop diagrams and
\begin{equation}
\de(V_2,V_2)=\frac{1}{(2\pi)^3}\int\,d^{\,3}p''\frac{V_2(\vec{p}{\;'},\vec{p}{\,''}) V_2(\vec{p}{\;''},\vec{p}\,)}{E_1(\vec{p}\,)+E_2(\vec{p}\,)-E_1(\vec{p}{\;''})- E_2(\vec{p}{\;''})} \,.
\end{equation}
The subtraction is necessary to prevent double counting of the Coulomb exchange in the box diagram.

Using our formulation, the $\al^2/r^2$ term in Eq.\,(\ref{GRS3}) arises from the $\de(V_2,V_2)$ subtraction term in Eq.\,(\ref{oneloop}). In momentum space this term behaves as $|\vec{k}|^{-1}$. As can be seen in Eqs.\,(2.3-2.7) of Ref.\,\cite{GRS}, the only term of this type that survives is the one in $\de(V_2,V_2)$, with $V_2(\vec{p}\;',\vec{p})$ given by
\begin{equation}
V_2(\vec{p}\;',\vec{p})=\frac{-e^2}{(\vec{p}\;'-\vec{p})^2+\la^2} \left(1-\frac{(\vec{p}\;'-\vec{p})^2}{8\mu^2}+\frac{\vec{p}^{\;2}}{m_1m_2}+ \cdots\right)\,,
\end{equation}
where the dots denote spin-dependent terms that are not relevant.

If one calculates the one-loop effective potential using the Breit-Pauli equation as a starting point, the corresponding subtraction term will involve a $V_2(\vec{p}\;',\vec{p})$ of the form
\begin{eqnarray}
V_2(\vec{p}\;',\vec{p})&=&\frac{-e^2}{(\vec{p}\;'-\vec{p})^2+\la^2} \left(1-\frac{1}{8}\Big(\frac{1}{m_1^2}+\frac{1}{m_2^2}\Big) (\vec{p}\;'-\vec{p})^2+ \frac{\vec{p}^{\;2}}{m_1m_2}\right. \nonumber \\ [6pt]
&& \left. +\frac{\vec{p}\Dot(\vec{p}\;'-\vec{p})\vec{p}\Dot(\vec{p}\;'-\vec{p})} {m_1m_2(\vec{p}\;'-\vec{p})^2}+\cdots\right)\label{V2BP}\,.
\end{eqnarray}

In this case, the last term in Eq.\,(\ref{V2BP}) exactly cancels the coefficient of the $|\vec{k}|^{-1}$ term produced by the second and third terms when $\de(V_2,V_2)$ is evaluated. Consequently, there is no $\al^2/r^2$ term in the one-loop corrections to the Breit-Pauli potential.

\end{document}